\documentclass[10pt,twoside,russian]{winter}
\RequirePackage[cp1251]{inputenc}
\RequirePackage[T2A]{fontenc}
\RequirePackage[russian,english]{babel} \RequirePackage{epsfig}
\RequirePackage{amsmath} \RequirePackage{amssymb}
\RequirePackage{amsfonts} \RequirePackage{longtable}
\usepackage{chapterbib}
\usepackage{indentfirst}
\usepackage{cite}

\usepackage{lscape}

\begingroup
\catcode`\+\active\gdef+{\mathchar8235\nobreak\discretionary{}%
 {\usefont{OT1}{cmr}{m}{n}\char43}{}}
\catcode`\-\active\gdef-{\mathchar8704\nobreak\discretionary{}%
 {\usefont{OMS}{cmsy}{m}{n}\char0}{}}
\catcode`\=\active\gdef={\mathchar12349\nobreak\discretionary{}%
 {\usefont{OT1}{cmr}{m}{n}\char61}{}}
\endgroup

\def\times{\mathchar8706\nobreak\discretionary{}%
 {\usefont{OMS}{cmsy}{m}{n}\char2}{}}
\AtBeginDocument{\mathcode`\==32768 \mathcode`\+=32768
\mathcode`\-=32768 }

\def\eqalign#1{\null\,\vcenter{\openup\jot\m@th
  \ialign{\strut\hfil$\displaystyle{##}$&$\displaystyle{{}##}$\hfil
      \crcr#1\crcr}}\,}
\def\eqalignleft#1{\null\,\vcenter{\openup\jot\m@th
  \ialign{\strut$\displaystyle{##}$\hfil&$\displaystyle{{}##}$\hfil
      \crcr#1\crcr}}\,}

\def\b#1:{{\bf#1}, }

\def\beq#1{\begin{equation}\label{#1}}
\def\eeq{\end{equation}}

\def\b{\begin{eqnarray}}
\def\earr{\end{eqnarray}}

\def\fun#1#2{\lower3.6pt\vbox{\baselineskip0pt\lineskip.9pt
\ialign{$\mathsurround=0pt#1\hfil##\hfil$\crcr#2\crcr\sim\crcr}}}

\def\b{\begin{eqnarray}}
\def\earr{\end{eqnarray}}

\def\fun#1#2{\lower3.6pt\vbox{\baselineskip0pt\lineskip.9pt
\ialign{$\mathsurround=0pt#1\hfil##\hfil$\crcr#2\crcr\sim\crcr}}}

\makeatletter
\renewcommand{\footnoterule}{\kern-3\p@
 \hrule width .4\columnwidth
 \kern 2.6\p@}
\renewcommand{\@makefnmark}{}
\renewcommand{\@makefntext}[1]{\hspace{2em}\hbox{\hss#1}}

\addto\captionsrussian{%
\def\abstractname{}
}

\renewcommand{\@biblabel}[1]{#1.}
\makeatother

\newcommand{\pud}{\hbox to 0.7em {\hspace{0.2em}.\hss $^d$}}
\newcommand{\pus}{\hbox to 0.4em {\hspace{0.01em}$''$\hss .}}
\newcommand{\pum}{\hbox to 0.7em {\hspace{0.2em}.\hss $^m$}}
\newcommand{\pug}{\hbox to 0.4em {\hspace{0.05em}.\hss $^{\circ}$}}
\newcommand{\Rim}[1]{\uppercase\expandafter{\romannumeral#1}}

\newcommand{\bnl}[1]{\begin{equation}\label{#1}}
\newcommand{\ed}{\end{equation}}

\begin{document}

\selectlanguage{russian}

\par
\hspace{0.3\textwidth}
\parbox[t]{0.65\textwidth}
{
\raggedleft
{\bf П.~Э.~Боли}\\
Институт астрономии общества Макса Планка (MPIA), Германия\\
\raggedright }

\footnote{\copyright\ П.~Э.~Боли, 2012}
\par
\medskip
\begin{center}
\bf
ЕВРОПЕЙСКАЯ ЮЖНАЯ ОБСЕРВАТОРИЯ И ТЕЛЕСКОПЫ VLT НА ПАРАНАЛЕ
\end{center}

\addcontentsline{toc}{subsection}{{\bf Боли П.~Э.} Европейская южная
обсерватория и телескопы VLT на Паранале}
\par
\medskip

\begin{minipage}{0.9\textwidth}
{\small Одним из важнейших в мире наблюдательных комплексов являются
телескопы Very Large Telescope (VLT) Европейской южной обсерватории
(ESO) на Серро Параналь.  В данной обзорной лекции даётся
характеристика организации ESO и телескопов VLT и проводится небольшая
выборка научных работ.  Особое внимание уделяется вопросу важности
телескопов VLT для российских учёных и перспективам использования
данных из ESO и VLT сегодня и в будущем.
}\end{minipage} \par \vspace{2mm}

\selectlanguage{english}

\begin{minipage}{0.9\textwidth}
{\small The Very Large Telescope (VLT) of the European Southern
Observatory (ESO) on Cerro Paranal is one of the most influential
observing complexes in the world.  In this overview lecture, an
introduction to the ESO organization and VLT telescopes is presented,
along with a small selection of scientific works.  Particular
attention is given to the importance of the VLT for the Russian
scientific community, as well as to present and future perspectives
for making use of ESO and VLT data.}\end{minipage}

\selectlanguage{russian}

\section*{Введение}

Определяющим инструментом для космических исследований всегда являлась
астрономическая обсерватория.  Первые, простые с технической точки
зрения обсерватории были построены отдельными исследователями или
небольшими группами людей.  Позже такой задачей, ввиду усложнения
требований к точности и чувствительности приборов, могли заниматься
только институты или университеты или даже группы, состоящие из
нескольких институтов.  В наши дни это зачастую дело международное, к
которому присоединяются не только отдельные единицы-институты, но даже
целые национальные академии и министерства.

Самой большой организацией, занимающейся построением телескопов и
обсерваторий, а также их управлением, является Европейская южная
обсерватория ESO (European Southern Observatory, полное название~---
European Organization for Astronomical Research in the Southern
Hemisphere), созданная в 1962~г.  В данный момент в неё входят 15
стран~--- Австрия, Бельгия, Бразилия, Великобритания, Германия, Дания,
Испания, Италия, Нидерланды, Португалия, Финляндия, Франция, Чехия,
Швейцария и Швеция.  Бюджет ESO составлял 163,2 млн евро в 2010~г., а
число рабочих и сотрудников~--- около 730 человек \cite{ESOAP10}.

Как правило, страны ESO участвуют в организации на условиях оплаты
вступительных и членских взносов, но есть также другие оговорённые
условия (например, разработка детекторов или оптических элементов).
Членство в ESO позволяет учёным соответствующих стран и институтов
бесплатно публиковать свои работы в рецензируемом журнале Astronomy
and Astrophysics и получать полную финансовую поддержку при проведении
наблюдений в обсерваториях организации.  Другими словами, все
астрономы институтов-членов ESO, от студентов до профессоров, могут
(при принятой заявке!) бесплатно приезжать в обсерватории в Чили,
проводить свои наблюдения и публиковать их результаты в признанном
журнале без каких-либо грантов или отчётов, что для обычного учёного,
конечно, очень удобно.

Данная обзорная лекция посвящена лишь одному из наблюдательных
комплексов в чилийской пустыне Атакама, которыми управляет ESO:
телескопам VLT на Серро Параналь (высота 2\,600~м).  Помимо телескопов
VLT, на Паранале также расположены обзорные телескопы VISTA и VST.
Кроме Паранальской обсерватории в список инструментов ESO входят
несколько оптических телескопов на пике Ла-Силья (высота 2\,400~м) и
12-м субмиллиметровый телескоп APEX на пике Чайнантор (высота
5\,100~м).  И наконец, ESO является партнёром в международном проекте
ALMA, который будет представлять собой интерферометр из 66
радиотелескопов (в настоящее время предварительные научные наблюдения
уже начались и достроена примерно одна треть всего массива).

\section*{Телескопы VLT}
\subsection*{Общие сведения}

Как уже отмечалось, телескопы VLT расположены на пике Параналь на
высоте 2\,600~м.  Сам пик Параналь находится на
$24^\circ38^\prime$~ю.~ш. $70^\circ24^\prime$~в.~д., примерно в 12~км
от побережья Тихого океана.  Здесь сверхнизкий уровень атмосферных
осадков (меньше 10~мм в год), и по виду это место очень напоминает
планету Марс.  Такие погодные условия способствуют наблюдениям на всех
длинах волн, особенно в инфракрасном диапазоне; устойчивость атмосферы
позволяет \emph{регулярно} достигать качества изображения до
0,7$^{\prime\prime}$ и иногда лучше 0,5$^{\prime\prime}$.  Кроме того,
расположение обсерватории в южном полушарии очень удобно для
наблюдения источников в направлении внутренней части Галактики.

VLT, несмотря на своё название (Very Large Telescope), на самом деле
состоит не из одного, а из восьми оптических телескопов: четырёх
неподвижных диаметром 8,2~м и четырёх подвижных диаметром 1,8~м.
Большие телескопы называются UT1, UT2, UT3, UT4 (UT~--- Unit
Telescope), а маленькие~--- AT1, AT2, AT3, AT4 (AT~--- Auxiliary
Telescope).  Первый 8-м телескоп начал свою работу в 1998~г.,
последний~--- в 2000~г.  Все телескопы UT оснащены полным комплектом
научных приборов и большую часть времени работают независимо друг от
друга, хотя телескопы AT специально предназначены для функционирования
только в интерферометрическом режиме и имеют оборудование лишь для
этой конкретной задачи.

\begin{figure}[t]
\centering
\includegraphics[width=0.8\textwidth]{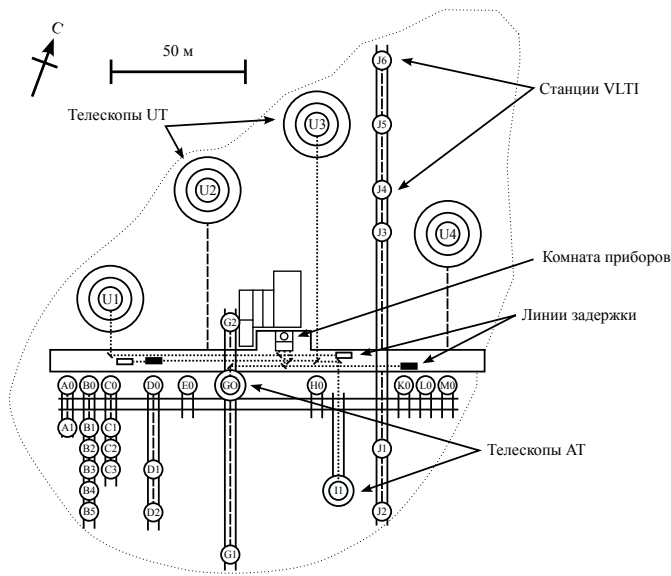}
\caption{Схема расстановки телескопов и интерферометрических станций
на VLT.  В данном виде два телескопа AT стоят на станциях G0 и I1.
Точечной линией показан путь оптических лучей от двух 1,8-м телескопов
AT и двух 8,2-м телескопов UT1, UT3, которые сходятся в комнате
приборов для наблюдения в интерферометрическом режиме.}
\label{sitelayout}
\end{figure}

На рисунке показано расположение телескопов и интерферометрических
станций на Паранале.  Для работы в интерферометрическом режиме
подземная система оптических туннелей позволяет направлять свет из
телескопов UT и 30 станций для телескопов AT в одну точку в комнате
приборов.  Подвижные телескопы AT располагаются по этим станциям в
зависимости от требований покрытия фазовой плоскости $uv$.  При
обычном порядке работы, техники два раза в неделю перемещают один из
телескопов по очереди, таким образом постепенно меняя конфигурацию.

\section*{Приборы и наблюдательные режимы}

Полный список приборов организации ESO, установленных на телескопах
VLT в текущее время, перечислен в таблице.  Кроме этих основных
(обсерваторских) инструментов астрономы также могут привести свою
аппаратуру для временного подключения к местным системам.

\begin{table}[t]
\caption{Наблюдательные приборы на телескопах VLT}
\label{instruments}
\begin{center}
\scriptsize
\begin{tabular}{lcccc}
\hline
\hline
Название & Телескоп & Режимы      & Спектральный & Спектральное \\
прибора  &          & наблюдения & диапазон     & разрешение\\
\hline
CRIRES & UT1 & Спектроскопия & 0,92-5,2 мкм & До 100\,000 \\
FLAMES & UT2 & Многообъекнтая & 370-950 нм & 5600-46\,000 \\
       &     & спектроскопия  &            & \\
FORS2  & UT1 & Прямые снимки, & 330-1100 нм & 100-400 \\
       &     & поляриметрия,  &             & \\
       &     & спектроскопия  &             & \\
HAWK-I & UT4 & Прямые снимки  & 0,85-2,5 мкм& --- \\
ISAAC  & UT3 & Прямые снимки, & 1-5 мкм & 200-10\,000 \\
       &     & спектроскопия  &         & \\
NACO   & UT4 & Прямые снимки, & 0,45-2,5 мкм & 400-1100 \\
       &     & поляриметрия,  &              & \\
       &     & коронография,  &              & \\
       &     & спекл- & & \\
       &     &интерферометрия & & \\
SINFONI& UT4 & Спектроскопия  & 1,1-2,45 мкм & 1500-4000 \\
       &     & (интегральное  &              & \\
       &     & поле)          &              & \\
UVES   & UT2 & Спектроскопия  & 300-1100 нм & До 110\,000 \\
VIMOS  & UT2 & Прямые снимки, & 360-1000 нм & 180-2500 \\
       &     & спектроскопия  & & \\
       &     & (многообъектная, & & \\
       &     & интегральное & & \\ 
       &     & поле) & & \\
VISIR  & UT3 & Прямые снимки, & 8-24,5 мкм & До 25\,000 \\
       &     & спектроскопия & & \\
XSHOOTER&UT2 & Спектроскопия & 300-2500 нм & До 14\,000 \\
\\
AMBER & VLTI & Интерферометрия & 1,05-2,4 мкм & До 12\,000 \\
MIDI  & VLTI & Интерферометрия & 8-13 мкм & 30-230 \\
\hline
\end{tabular}
\end{center}
{\footnotesize {\it Примечания.} Спектральное разрешение определяется
как $\lambda / \Delta \lambda$.  Интерферометрические инструменты VLTI
(AMBER, MIDI) могут работать с любыми телескопами UT или AT.  В списке
приведены только общедоступные приборы.}
\end{table}

Как видно из таблицы, наблюдения проводятся в различных режимах
(включая прямые снимки, спектроскопию, интегральное поле,
поляриметрию, коронографию и интерферометрию) в диапазонах от ближнего
ультрафиолетового до среднего инфракрасного.  Соответственно список
реализуемых задач довольно широкий; в 2010~г. количество рецензируемых
публикаций на основе данных с VLT составило 507, а с 1999 по
2010~г.~--- 3\,669.  В число наиболее цитируемых работ входят
измерения замедления расширения Вселенной \cite{Riess04}, наблюдения
звёзд на орбитах вокруг сверхмассивной чёрной дыры в центре
Галактики \cite{Schoedel02}, изучение оптического (в покое) спектра
галактик типа лаймановского скачка \cite{Pettini01}, наблюдения звёзд
с очень низкой металличностью \cite{Cayrel04} и др.

\subsection*{Интерферометр VLTI}

Помимо стандартных режимов наблюдения одной из главных задач
телескопов VLT являются интерферометрические наблюдения.  Массив
телескопов, работающих в этом режиме, называется VLTI (Very Large
Telescope Interferometer).  В данном случае интерферометр представляет
собой объединение двух~--- четырёх телескопов VLT (включая как 8,2-м
телескопы UT, так и 1,8-м телескопы AT).  Пучок лучей от телескопов
направляется через систему зеркал в подземных туннелях и сходится в
комнате интерферометрических приборов (см. рисунок).  Там записывается
картина интерференции (в отличие от радиоинтерферометров с
гетеродинными детекторами, где интерференционный узор строится не в
реальном времени), из которой определяется коррелированный поток во
время обработки данных.  Таким образом извлекается сигнал,
соответствующий шкалам $\lambda / 2 B$, где $\lambda$~--- длина волны;
$B$~--- расстояние между двумя телескопами, составляющими
интерферометр.  На длине волны 10 мкм при базисе 100~м, например,
достигнутое пространственное разрешение составляет
0,01$^{\prime\prime}$ (10~а.~э. на расстояние 1~кпк).

\section*{Наблюдения на VLT}
\subsection*{Заявки}

Чтобы получить возможность наблюдения на любых телескопах VLT, нужно
подать заявку на наблюдательное время.  ESO делит наблюдения на два
периода, которые начинаются в апреле и в октябре.  На эти полугодия
объявляются конкурсы, заявки принимаются до конца сентября и марта
соответственно.  На каждый период подаются около 2\,000 заявок, что
превышает общее наблюдательное время примерно в три раза.  Кроме того,
для срочных программ в рамках резервного времени директора заявки
принимаются на рассмотрение в любое время года.

Заявки ESO должны быть лаконичными и содержать чёткое научное
обоснование в пределах двух страниц. В заявке должны быть подробно
описаны количественные оценки суммарного наблюдательного времени,
ожидаемых уровней сигнала и т.~п.  Заявки рассматривает международный
комитет, в который входят 15--20 учёных из разных институтов.  Задача
комитета состоит в том, чтобы отбирать лучшие и более перспективные
научные программы, оценивая их актуальность и вероятность успешного
выполнения на ряду с другими предложенными проектами.

\subsection*{Наблюдения в обычном и удалённом режимах}

Если заявки принята, наблюдения проводятся как в удалённом режиме, так
и в обычном (сам астроном ездит на обсерваторию) в зависимости от
сложности задачи.  В настоящее время примерно 60--70\% от общего
количества наблюдений проводится в удалённом режиме.  Для работы в
данном режиме наблюдательная программа составляется заранее, за
несколько недель, в виде специального запроса со всеми настройками
приборов и нужной информацией для персонала обсерватории.  Этот заказ
на наблюдения ставится в очередь в соответствии с установленным
комитетом приоритетом (высокий, средний или низкий) и исполняется,
когда условия (погода, часовой угол, отсутствие других программ с
более высокими приоритетами) позволяют.  Сами наблюдения проводятся
паранальскими астрономами без участия заявителя, наблюдательные данные
потом можно скачать в необработанном виде из архива через Интернет.
Для некоторых инструментов ESO посылает автоматически обработанные
данные заявителю проекта к концу наблюдательного полугодия.

При обычном режиме наблюдатель сам ездит на Параналь.  К данному
режиму относятся сложные или нестандартные наблюдения или те, риск
неправильного выполнения которых ESO не хочет брать на себя.
Добираться до обсерватории довольно долго: для работающего в Европе
астронома, например, поездка начинается с пятнадцатичасового рейса
через шесть часовых поясов до столицы Чили Сантьяго.  Там ему
предоставляется ночлег в доме для гостей ESO, блюда местной кухни,
приготовленные профессиональным поваром, и, по обычаю, чилийский
национальный напиток «писко сауер».  Другими словами, организация ESO
делает всё, чтобы после тяжёлой поездки и перед долгими ночами за
телескопом астрономам было максимально комфортно.

Проведя сутки-двое в Сантьяго, астроном продолжает свой путь до
обсерватории, которая находится в 1\,000~км на север от чилийской
столицы.  По правилам ESO астрономам положено приезжать на
обсерваторию за день-два до начала наблюдений.  Как и в Сантьяго,
условия быта для астрономов и инженеров на горе очень благоприятные.
На высоте 2\,000~м, в 3~км от пика Параналь, расположена удостоенная
наград гостиница ESO.  Здание встроено прямо в землю.  Это частично
замкнутая экологическая система, чтобы длительно находящиеся на горе
люди не страдали из-за экстремальных погодных условий (относительная
влажность обычно составляет 5-15\%), которые совсем непригодны для
жизни (так, в 2003~г. был проведён эксперимент, в рамках которого
искали следы жизни в почве пустыни Атакама с использованием
применённых на спускаемых аппаратах «Викинг» методов.  Следов жизни на
основе ДНК найдено не было \cite{NG03}).

Во время наблюдений астроном работает в общем зале управления, в
котором располагаются все терминалы, оборудование и персонал для
управления телескопами UT, VLTI, VST и VISTA (последние два телескопа
также находятся на Паранале, но не относятся к VLT).  Телескопом
управляют как минимум двое~--- инженер, который обеспечивает
нормальную работу телескопа и научных приборов и в то же время
старается защитить дорогостоящее оборудование от небрежного обращения,
и обсерваторский астроном, который выполняет указы и просьбы приезжего
наблюдателя.  Внеобсерваторским людям (т.~е. нам) самим управлять
телескопами строго запрещено.  После наблюдений полученный материал
записывается на носители или передаётся по Интернету, и астроном
начинает долгий путь обратно, хотя некоторые посвящают недельку-две
отдыху в Чили.

\subsection*{Архив наблюдательных данных ESO}

Все данные, полученные на телескопах VLT (и на всех телескопах ESO),
сохраняются в архиве, который доступен в Интернете по адресу:
\texttt{http://archive.eso.org}.  Доступ к наблюдательным данным 
того или иного проекта представлен исключительно заявителю программы в
течение одного года после наблюдений.  После этого срока данные
выходят в открытый доступ и их может скачать каждый желающий (не
только астрономы из стран ESO).  Общий размер материалов наблюдений
материала в архиве в данный момент составляет примерно 65~ТБ и
увеличивается на 15~ТБ в год.

Открытый доступ к наблюдательным данным даёт возможность не только
проверять уже проделанную работу, но и использовать эти данные в своих
собственных проектах.  Например, как и на всех обсерваториях, очень
часто получается так, что данные были получены, но не анализировались
или что старые наблюдения можно использовать в новом качестве.  Такой
подход относится к идее так называемой виртуальной обсерватории, о
которой можно более подробно прочитать в трудах 37-й зимней
школы \cite{Malkov08}.

\section*{Примеры работы российских астрономов на VLT}

Несмотря на то что Россия пока не является членом ESO, несколько
российских астрономов уже работают с полученным на VLT наблюдательным
материалом.  Здесь приведена небольшая выборка из трёх работ,
написанных российскими астрономами на основе данных из VLT.
\begin{itemize}
\item Т.~А.~Рябчикова и др. представили работу в
2007~г. В ней исследуются вертикальные моды колебаний быстро
колеблющихся звёзд типа Ap (roAp) \cite{Ryabchikova07}.  Использованы
958 спектров в оптическом диапазоне, полученных на спектрографе UVES,
для выборки~--- восемь звёзд типа roAp.  Посредством исследования
кратковременных вариаций большого набора различных спектральных линий
авторам удалось раскрыть вертикальную структуру мод колебаний и
химической стратификации в звёздных атмосферах.

\item В 2010~г. в своей магистерской диссертации 
М.~С.~Храмцова изучала поглощающие системы на красном смещении
$z\sim0.4$ по лучу зрения квазаров \cite{Khramtsova10}.  В данной
работе использован прибор VIMOS в режиме интегрального поля, что
позволяет получить 1\,600 спектров в поле зрения $54 \times
54^{\prime\prime}$.  Исследуется характер поглощающего вещества, и в
некоторых случаях эти системы отождествляются с образующими звёзды
галактиками.

\item В работе 2011~г., которая заняла первое место на студенческом
конкурсе 40-й зимней школы, Т.~М.~Ситнова и Л.~И.~Машонкина
проанализировали вклад $r$- («быстрого») и $s$- («медленного»)
процессов захвата нейтронов в химический состав звезды гало
HD~29907 \cite{Sitnova11}.  В работе используются спектры из архива,
снятые на спектрографе UVES.  Тщательный анализ обилия тяжёлых
элементов позволил сделать заключения об условиях межзвёздной среды во
время формирования звезды и о ролях различных механизмов в
нуклеосинтезе.
\end{itemize}

\section*{Россия, ESO и Вы}

В последнее время разговоры о возможном вступлении России в ESO идут
на высоком уровне (см., например, \cite{itartass11}).  Хотя такие
обсуждения выходят за рамки данной лекции, важно отметить, что
членство в ESO дало бы российским астрономам всех уровней огромную
выгоду.  Не исключено, например, что первокурсники этой же зимней
школы смогут уже ко времени своей магистерской диссертации или
дипломной работы написать собственную заявку на наблюдения в ESO,
провести наблюдения в Чили и опубликовать результаты в журнале
Astronomy and Astrophysics без каких-либо специальных грантов.  Однако
для членства в ESO на таких прекрасных условиях России придётся
платить~--- примерно 10-15~млн евро (400--600~млн российских рублей) в
год.

Тем не менее и сейчас мотивированному студенту ничего не мешает
связаться со своими коллегами из других стран и подать заявку на
наблюдения в ESO.  Опыт свидетельствует о том, что европейские институты
приветствуют такой подход и очень часто готовы поддержать работу с
иностранными коллегами за счёт своих средств.  И это значит, что все
перспективы для сотрудничества уже есть.

\section*{Заключение}

Телескопы VLT на Паранальской обсерватории в Чили являются ключевым
инструментом Европейской южной обсерватории.  Широкий набор
наблюдательных приборов в диапазоне от 300~нм до 25~мкм обеспечивает
большое количество решаемых задач, начиная с исследований объектов в
нашей Солнечной системе и заканчивая дальними галактиками и квазарами.
Поддержка со стороны ESO делает весь процесс максимально удобным и
доступным для астрономов, способствуя таким образом научному
прогрессу.

Конкурс для получения наблюдательного времени на телескопах VLT
довольно большой, и время преимущественно даётся астрономам из
стран--членов ESO.  Однако доступ к архиву наблюдательных данных ESO
предоставляется всем астрономам мира и является бесценным ресурсом,
особенно для тех, кому тяжело или невозможно получить время на
подобных телескопах.  Поэтому, несмотря на то, что Россия не является
членом ESO, возможность использовать её наблюдения всё равно есть, и
этим нужно воспользоваться.

Пока невелико количество работ, выполняемых астрономами из России с
наблюдательным материалом из VLT, но несколько групп работают с
ресурсами паранальской обсерватории уже с момента её открытия.  В
список публикуемых статей входят работы как студентов и аспирантов,
так и сотрудников научных институтов.  И в конечном итоге совершенно
справедливо ожидать, что количество их будет только расти.


\def\apj{Astrophys.~J}
\def\aatr{Astron.~Astroph.~Trans}
\def\aaps{Astron.~and Astrophys.~Suppl.~Ser}
\def\pasp{Publ.~Astron.~Soc.~Pac}
\def\gca{Geochim.~Cosmochim.~Acta}
\def\aap{Astron.~Astrophys}
\def\aspcs{ASP~Conf.~Ser}
\def\asrep{Astron.~Rep}
\def\nat{Nature}
\def\apjl{Astrophys.~J.,~Lett}
\def\apjs{Astrophys.~J.,~Suppl.~Ser}
\def\aj{Astron.~J}
\def\mnras{Mon.~Not.~R.~Astron.~Soc}
\def\araa{Ann.~Rev.~Astron.~Astrophys}
\def\jcp{J.~Chem.~Phys}
\def\apss{Astrophys.~Space.~Sci}
\def\prl{Phys.~Rev.~Lett}
\def\phrva{Phys.~Rev.~A}
\def\phlb{Phys.~Let.~B}
\def\pf{Phys.~Fluids}
\def\azh{Астрон.~журн}
\def\pazh{Письма~в~Астрон.~журн}
\def\jgr{J.~Geophys.~Res}
\def\cemda{Celest.~Mech.~Dyn.~Astr}
\def\jcoph{J.~Comp.~Phys}
\def\cophc{Comput.~Phys.~Commun}
\def\phpl{Physics~of~Plasmas}
\def\pasj{Publ.~Astron.~Soc.~Jpn}
\def\avest{Астрон.~вест}
\def\jrasc{J.~R.~Astron.~Soc.~Can}
\def\cemec{Celest.~Mech}
\def\pasau{Proc.~Astron.~Soc.~Aust}
\def\puasau{Publ.~Astron.~Soc.~Aust}
\def\jasa{J.~Acoust.~Soc.~Am}
\def\jfm{J.~Fluid~Mech}
\def\cajph{Can.~J.~Phys}
\def\mitag{Mitt.~Astron.~Ges}
\def\bain{Bull.~Astron.~Inst.~Neth}
\def\epsl{Earth~Planet.~Sci.~Lett}
\def\ibvs{Inf.~Bull.~Variable~Stars}
\def\arep{Astr.~Rep}
\def\phr{Phys.~Rep}
\def\astl{Astron.~Letters}
\def\sci{Science}
\def\jqsrt{J.~Quant.~Spectrosc.~Radiat.~Transfer}
\def\emp{Earth,~Moon~and~Planets}
\def\icar{Icarus}
\def\pss{Planet.~Space~Sci}
\def\qjras{Q.~J.~R.~Astron.~Soc}
\def\nimpa{Nucl.~Instrum.~Methods~Phys.~Res.,~Sect.~A}
\def\soph{Sol.~Phys}
\def\lnm{Lect.~Notes~in~Math}
\def\an{Astron.~Nach}
\def\aph{Astroparticle~Physics}
\def\adspr{Adv.~Space~Res}
\def\geoj{Geophys.~J}
\def\caosp{Contrib.~Astron.~Obs.~Skalnat{\'e}~Pleso}
\def\vestcpbu{Вестн.~С.-Петерб.~ун-та}
\def\izvvrad{Изв.~вузов.~Радиофизика}
\def\izvans{Изв.~AH~CCCP}
\def\vestvgu{Вестн.~ВолГУ}
\def\bamass{Bull.~Am.~Astron.~Soc}
\def\rmxaa{Rev.~Mex.~Astron.~Astrofis}
\def\aapr{Astron.~Astrophys.~Rev}
\def\acp{Atmosphere~Chem.~Phys}
\def\cosiss{Космич.~исслед}
\def\ssrv{Space~Sci.~Rev}
\def\jmph{J.~Math.~Phys}
\def\rvmps{Rev.~Mod.~Phys.~Suppl}
\def\rvmp{Rev.~Mod.~Phys}
\def\prd{Phys.~Rev.~D}
\def\nuphs{Nuc.~Phys.~B~Proc.~Suppl}
\def\nuphb{Nuc.~Phys.~B}
\def\skytel{Sky~Telesc}
\def\thmc{Тез.~международ.~конф.}
\def\mmc{Материалы~международ.~конф.}
\def\mvrc{Материалы~всерос.~конф.}
\def\cntc{Сб.~науч.~тр.~конф.}
\def\tmnpc{Тр.~Международ.~науч.-практ.~конф.}
\def\ctc{Сб.~тр.~конф.}
\def\mcnc31{Тр.~31-й~Международ.~студ.~науч.~конф., Екатеринбург, 28 янв.---1 февр. 2002~г}
\def\mcnc32{Тр.~32-й~Международ.~студ.~науч.~конф., Екатеринбург, 3---7 февр. 2003~г}
\def\mcnc33{Тр.~33-й~Международ.~студ.~науч.~конф., Екатеринбург, 2---6 февр. 2004~г}
\def\mcnc34{Тр.~34-й~Международ.~студ.~науч.~конф., Екатеринбург, 31 янв.---4 февр. 2005~г}
\def\mcnc35{Тр.~35-й~Международ.~студ.~науч.~конф., Екатеринбург, 30 янв.---3 февр. 2006~г}
\def\mcnc36{Тр.~36-й~Международ.~студ.~науч.~конф., Екатеринбург, 29 янв.---2 февр. 2007~г}
\def\mcnc37{Тр.~37-й~Международ.~студ.~науч.~конф., Екатеринбург, 28 янв.---1 февр. 2008~г}
\def\mcnc38{Тр.~38-й~Международ.~студ.~науч.~конф., Екатеринбург, 2---6 февр. 2009~г}
\def\mcnc39{Тр.~39-й~Международ.~студ.~науч.~конф., Екатеринбург, 1---5 февр. 2010~г}
\def\mcnc40{Тр.~40-й~Международ.~студ.~науч.~конф., Екатеринбург, 31 янв.---4 февр. 2011~г}
\def\tmc{Тр.~Международ.~конф.}
\def\tc{Тр.~конф.}
\def\thc{Тез.~конф.}

\def\IAUsymp{Proc.~IAU~Symp.}
\def\IAUcoll{Proc.~IAU~Colloquia}
\def\prconf{Proc.~conf.}
\def\princonf{Proc.~int.~conf.}
\def\mvrnc{Материалы~всерос.~науч.~конф}

\bibliography{boley}

\end{document}